# Off–axis Beams and Detector Clusters: Resolving Neutrino Parameter Degeneracies

V. Barger[1], D. Marfatia[2] and K. Whisnant[3]

[1]*Department of Physics, University of Wisconsin, Madison, WI 53706, USA*
[2]*Department of Physics, Boston University, Boston, MA 02215, USA*
[3]*Department of Physics and Astronomy, Iowa State University, Ames, IA 50011, USA*

## Abstract

There are three parameter degeneracies inherent in the three–neutrino analysis of long–baseline neutrino experiments. We develop a systematic method for determining whether or not a set of measurements in neutrino oscillation appearance experiments with approximately monoenergetic beams can completely resolve these ambiguities. We then use this method to identify experimental scenarios in which the parameter degeneracies may be efficiently resolved. Generally speaking, with two appearance measurements degeneracies can occur over wide areas of the $(\delta, \theta_{13})$ parameter space; with three measurements they occur along lines in the parameter space and with four measurements they occur only at isolated points. If two detectors are placed at the same distance from the source but at different locations with respect to the main axis of the beam (a detector cluster), each detector will measure neutrinos at different energies. Then one run with neutrinos and one run with antineutrinos will give the four independent measurements that in principle can resolve all of the parameter degeneracies if $\sin^2 2\theta_{13} \geq 0.002$. We also examine scenarios with detector clusters using only neutrino beams. Without detector clusters, the measurement of neutrinos and antineutrinos at a short distance and only neutrinos at a longer distance may also work.

# I. INTRODUCTION

The recent solar neutrino data from the Sudbury Neutrino Observatory [1,2], which infers different neutrino fluxes from the charged– and neutral–current measurements, provides convincing evidence that electron neutrinos do in fact change flavor as they travel from the Sun to the Earth. The neutral–current measurement [2] is also consistent with the solar neutrino flux predicted in the Standard Solar Model [3]. The preponderance of solar neutrino data increasingly prefers the Large Mixing Angle (LMA) solution to the solar neutrino puzzle, with $\delta m_{21}^2 \sim 5 \times 10^{-5}$ eV$^2$ and amplitude close to 0.8 [2,4,5]. This solution will be tested decisively by the KamLAND reactor neutrino experiment [6].

The atmospheric neutrino deficit also gives a strong indication that neutrinos have mass and oscillate from one flavor to another – the most compelling interpretation is that $\nu_\mu$'s created in the atmosphere oscillate to $\nu_\tau$ with almost maximal amplitude and mass-squared difference $\delta m_{31}^2 \sim 3 \times 10^{-3}$ eV$^2$ [7]. The K2K experiment [8] with a baseline of 250 km has preliminary results that are in agreement with this interpretation. Oscillations of $\nu_\mu$ to $\nu_e$ as an explanation of the atmospheric anomaly are ruled out by the CHOOZ [9] and Palo Verde [10] reactor experiments, which place a bound on the $\nu_\mu \to \nu_e$ oscillation amplitude smaller than 0.1 at the 95% C.L. in the $\delta m_{31}^2$ region of interest. The MINOS [11], ICARUS [12] and OPERA [13] experiments are expected to come online in 2005 and study aspects of the oscillations at the atmospheric scale [14]. The low energy beam at MINOS will allow a very accurate determination of the leading oscillation parameters. ICARUS and OPERA should provide concrete evidence that $\nu_\mu \to \nu_\tau$ oscillations are responsible for the atmospheric neutrino deficit by identifying tau neutrino events. These long–baseline experiments, when combined with KamLAND, should provide further information on all of the parameters in the three–neutrino mixing matrix, except for the $CP$–violating phase ($\delta$) and possibly not the mixing angle associated with $\nu_\mu \to \nu_e$ oscillations in atmospheric and long–baseline experiments ($\theta_{13}$). It will take a new generation of experiments to provide accurate measurements of these parameters.

Future measurements of $\theta_{13}$ and $\delta$ are not completely straightforward. The parameter $\theta_{13}$ cannot be determined from a single $\nu$ measurement since the measured oscillation rate also depends on the value of $\delta$, which is not known. Even with both a $\nu$ oscillation measurement and a $\bar{\nu}$ oscillation measurement there are several parameter degeneracies that enter the determination of the three–neutrino mixing matrix. These include the ambiguities associated with (i) the parameters in the $U_{e3}$ (=$\sin\theta_{13}$ e$^{-i\delta}$) element [15–19], (ii) the sign of the mass–squared difference responsible for the oscillations of atmospheric neutrinos ($\delta m_{31}^2$) [16–20], and (iii) the primary mixing angle for the oscillation of atmospheric neutrinos ($\theta_{23}$) [17,19]. Each of these ambiguities can mix $CP$ violating ($CPV$) and $CP$ conserving ($CPC$) solutions. To resolve the ambiguities requires multiple measurements; the usual prescription involves making measurements with different beam energies, baseline lengths, and/or beam particles (neutrinos versus antineutrinos). A typical scenario requires measurements of neutrinos and antineutrinos, each at two different energies [17], which would entail four separate runs, each of which would take years to complete.

In this paper we discuss alternate possibilities that capitalize on the fact that the energy spectrum of a conventional neutrino beam varies with the angle from the main beam axis [21–23]. The off–axis components of a neutrino beam have very narrow spectra for a given off–axis angle. Viewed this way, a single neutrino beam is actually a continuum of approximately monoenergetic beams; the energy can be tuned simply by varying the angle with respect to the main beam axis. This has the advantage that experiments at more than one neutrino energy may be run simultaneously using a single beam line, simply by placing detectors at slightly different angles from the main axis of the beam (a detector cluster). One could also locate one detector on the beam axis and another off–axis. Furthermore, the high energy tail of off–axis components of the beam is suppressed, reducing the high–energy contribution to backgrounds. One disadvantage is that the neutrino flux decreases as the off–axis distance increases, which may give a practical limit for the choices of neutrino energies.

Detailed studies have already been made for off–axis beams with single detectors [19,24,25], but how to resolve the parameter ambiguities has not been thoroughly addressed. In this paper we present some possible scenarios for resolving parameter degeneracies using detector clusters with off–axis neutrino beams. In particular we examine how well parameter degeneracies may be resolved with one run of a neutrino beam and one run of an antineutrino beam. We also discuss other, more speculative, possibilities involving detector clusters, such as multiple detectors and only neutrino beams. We compare these results to the ability of more conventional experimental setups (without detector clusters) to resolve parameter degeneracies.

The organization of our paper is as follows. In Sec. II we present oscillation probability formulas based on a constant density approximation that works well for the situations that we consider. In Sec. III we determine in a general way the number and type of measurements that are needed to resolve parameter degeneracies. Specific detector scenarios are presented and discussed in Sec. IV. Concluding remarks are made in Sec. V.

## II. OSCILLATION PROBABILITIES IN MATTER

We work in the three–neutrino scenario. For oscillation studies, the neutrino mixing matrix $U$ can be specified by 3 mixing angles $(\theta_{23}, \theta_{12}, \theta_{13})$ and a $CP$-violating phase $\delta$. We adopt the parametrization

$$U = \begin{pmatrix} c_{13}c_{12} & c_{13}s_{12} & s_{13}e^{-i\delta} \\ -c_{23}s_{12} - s_{13}s_{23}c_{12}e^{i\delta} & c_{23}c_{12} - s_{13}s_{23}s_{12}e^{i\delta} & c_{13}s_{23} \\ s_{23}s_{12} - s_{13}c_{23}c_{12}e^{i\delta} & -s_{23}c_{12} - s_{13}c_{23}s_{12}e^{i\delta} & c_{13}c_{23} \end{pmatrix}, \qquad (1)$$

where $c_{jk} \equiv \cos\theta_{jk}$ and $s_{jk} \equiv \sin\theta_{jk}$. For Majorana neutrinos there are two additional phases, but they do not affect oscillations. In the most general $U$, the $\theta_{ij}$ are restricted to the first quadrant, $0 \leq \theta_{ij} \leq \pi/2$, with $\delta$ in the range $0 \leq \delta < 2\pi$. We assume that $\nu_3$ is the neutrino eigenstate that is separated from the other two, and that the sign of $\delta m_{31}^2$ can be either positive or negative, corresponding to the case where $\nu_3$ is either above or below, respectively, the other two mass eigenstates. The magnitude of $\delta m_{31}^2$ determines the oscillation length of atmospheric neutrinos, while the magnitude of $\delta m_{21}^2$ determines the oscillation length of solar neutrinos, and thus $|\delta m_{21}^2| \ll |\delta m_{31}^2|$. If we accept the likely conclusion that the solar solution is LMA [2,4,5], then $\delta m_{21}^2 > 0$ and we can restrict $\theta_{12}$ to the range $[0, \pi/4]$. It is known from reactor neutrino data that $\theta_{13}$ is small, with $\sin^2 2\theta_{13} \leq 0.1$ at the 95% C.L. [9]. Thus a set of parameters that unambiguously spans the space is $\delta m_{31}^2$ (magnitude and sign), $\delta m^2_{21}$, $\sin^2 2\theta_{12}$, $\sin \theta_{23}$, and $\sin^2 2\theta_{13}$; only the $\theta_{23}$ angle can be below or above $\pi/4$.

To determine oscillation probabilities we use the constant density approximation and expand in terms of the small parameters $\theta_{13}$ and $\delta m^2_{21}$ [26,27], which has been shown to reproduce well the exact oscillation probabilities for $E_\nu > 0.5$ GeV, $\theta_{13}$ not too large, and $L < 4000$ km [17]. Up to second order in $\alpha$ and $\theta_{13}$, the oscillation probabilities for $\delta m^2_{31} > 0$ and $\delta m^2_{21} > 0$ are

$$P(\nu_\mu \to \nu_e) = |xf + yge^{i(\Delta+\delta)}|^2 = x^2 f^2 + 2xyfg(\cos\delta\cos\Delta - \sin\delta\sin\Delta) + y^2 g^2, \quad (2)$$

$$\bar{P}(\bar{\nu}_\mu \to \bar{\nu}_e) = |x\bar{f} + yge^{i(\Delta-\delta)}|^2 = x^2 \bar{f}^2 + 2xy\bar{f}g(\cos\delta\cos\Delta + \sin\delta\sin\Delta) + y^2 g^2, \quad (3)$$

respectively, where

$$x \equiv \sin\theta_{23} \sin 2\theta_{13}, \quad (4)$$
$$y \equiv \alpha \cos\theta_{23} \sin 2\theta_{12}, \quad (5)$$
$$f, \bar{f} \equiv \sin((1 \mp \hat{A})\Delta)/(1 \mp \hat{A}), \quad (6)$$
$$g \equiv \sin(\hat{A}\Delta)/\hat{A}, \quad (7)$$

and

$$\Delta \equiv |\delta m^2_{31}|L/4E_\nu = 1.27|\delta m^2_{31}/\text{eV}^2|(L/\text{km})/(E_\nu/\text{GeV}), \quad (8)$$
$$\hat{A} \equiv |A/\delta m^2_{31}|, \quad (9)$$
$$\alpha \equiv |\delta m^2_{21}/\delta m^2_{31}|. \quad (10)$$

In Eq. (9), $A/2E_\nu$ is the amplitude for coherent forward charged-current $\nu_e$ scattering on electrons, with

$$A = 2\sqrt{2}\, G_F\, N_e\, E_\nu = 1.52 \times 10^{-4}\, \text{eV}^2 Y_e\, \rho\, (\text{g/cm}^3) E_\nu\, (\text{GeV}), \quad (11)$$

and $N_e$ is the electron number density, which is the product of the electron fraction $Y_e(x)$ and matter density $\rho(x)$. In the Earth's crust and mantle the average matter density is typically 3–5 g/cm$^3$ and $Y_e \simeq 0.5$. In all of our calculations we use the average $N_e$ along the neutrino path, assuming the Preliminary Reference Earth Model [28].

The coefficients $f$ and $\bar{f}$ differ due to matter effects ($\hat{A} \neq 0$). The values of $f$, $\bar{f}$, and $g$ (scaled by $\sqrt{E_\nu}/L$ to account for the dependence of neutrino cross section with energy and the flux with distance) are shown versus $\Delta$ in Fig. 1 for $L = 300$ km (JHF to Kamioka), 730 km (Fermilab to Soudan or CERN to Gran Sasso), 1290 km (Fermilab to Homestake), 1770 km (Fermilab to Carlsbad), and 2900 km (Fermilab to SLAC); the scaling factor is chosen this way since the oscillation probabilities are quadratic functions of $f$, $\bar{f}$, and $g$.

To obtain the probabilities for $\delta m^2_{31} < 0$, the transformations $\hat{A} \to -\hat{A}$, $y \to -y$ and $\Delta \to -\Delta$ (implying $f \leftrightarrow -\bar{f}$ and $g \to -g$) can be applied to the probabilities in Eqs. (2) and (3) to give

$$P(\nu_\mu \to \nu_e) = x^2 \bar{f}^2 - 2xy\bar{f}g(\cos\delta\cos\Delta + \sin\delta\sin\Delta) + y^2 g^2, \quad (12)$$
$$\bar{P}(\bar{\nu}_\mu \to \bar{\nu}_e) = x^2 f^2 - 2xyfg(\cos\delta\cos\Delta - \sin\delta\sin\Delta) + y^2 g^2. \quad (13)$$

In practice, neutrino beams are not monoenergetic, even for a narrow band beam or an off–axis component of a beam. The number of appearance events is $N(\nu_\mu \to \nu_e) = \int P(\nu_\mu \to \nu_e)\,\Phi\sigma\,dE_\nu$, where $\Phi$ is the energy–dependent neutrino flux, $\sigma$ the interaction cross section, and $P$ the oscillation probability from Eq. (2). Then for $\delta m_{31}^2 > 0$, $N$ has the approximate form

$$N(\nu_\mu \to \nu_e) = A_1 x^2 + A_2 x \cos\delta + A_3 x \sin\delta + A_4\,, \tag{14}$$

where

$$\begin{aligned}
A_1 &= \int f^2 \Phi\sigma\,dE_\nu\,, \\
A_2 &= \int 2yfg \cos\Delta\,\Phi\sigma\,dE_\nu\,, \\
A_3 &= -\int 2yfg \sin\Delta\,\Phi\sigma\,dE_\nu\,, \\
A_4 &= \int y^2 g^2 \Phi\sigma\,dE_\nu\,.
\end{aligned} \tag{15}$$

There are similar expressions for $\bar{N}(\bar{\nu}_\mu \to \bar{\nu}_e)$, and for $\delta m_{31}^2 < 0$. In all cases, whether one is describing the number of events at a single energy or total events integrated over an energy spectrum, the result can be written approximately as a linear combination of $x^2$, $x\cos\delta$, and $x\sin\delta$. In fact the probability for constant density can be written as $A\cos\delta + B\sin\delta + C$ without any approximation [29]. It is this generic property that we exploit in this paper, so many of the qualitative aspects of our analysis should hold for neutrino beams that are not monoenergetic if only the total number of events is being used.

### III. DETERMINING THE OSCILLATION PARAMETERS

#### A. Parameter degeneracies with two measurements

It is expected that $\delta m_{31}^2$ and $\sin^2 2\theta_{23}$ will be well–measured in $\nu_\mu$ survival experiments; if the solar solution is LMA, as it now seems to be [2,4,5], then $\delta m_{21}^2$ and $\sin^2 2\theta_{12}$ will be well–determined by KamLAND [6]. Thus it is the parameters $\theta_{13}$ and $\delta$ that will be primarily measured in appearance experiments, modulo uncertainties in the other parameters.

A single measurement is not sufficient to determine $\theta_{13}$ since the oscillation probability depends on both $\theta_{13}$ and $\delta$ (see, e.g., Eqs. (2) and (3)). In fact the inferred measurement of $\sin^2 2\theta_{13}$ could be uncertain by an order of magnitude with only one measurement [17]. The usual approach for determining $\theta_{13}$ and $\delta$ is a measurement of both $P$ and $\bar{P}$ at one $L$ and $E_\nu$. Then $\theta_{13}$ and $\delta$ can be determined from Eqs. (2) and (3). Unfortunately, there are usually two solutions with distinct $\theta_{13}$ and $\delta$ (the well–known $(\delta, \theta_{13})$ ambiguity [15–19]). The existence of this ambiguity can be understood as follows. If $\theta_{13}$ is held fixed, then as $\delta$ is varied an ellipse is traced out in $(P, \bar{P})$ space; the eccentricity of the ellipse is determined by the oscillation argument $\Delta$ [17,20]. Ellipses for two different values of $\theta_{13}$ can intersect, so that a single point in $(P, \bar{P})$ space can be obtained by two different sets of parameters $(\delta, \theta_{13})$. For a given $(P, \bar{P})$ there are also often two solutions for $\delta m_{31}^2 < 0$ from Eqs. (12) and (13) (the $\mathrm{sgn}(\delta m_{31}^2)$ ambiguity). Furthermore, since only $\sin^2 2\theta_{23}$ is known from atmospheric neutrino experiments (or from measuring $\nu_\mu$ survival), if $\theta_{23} \neq \pi/4$, each solution discussed above also has two possible values of $\theta_{23}$ (each with a different $\theta_{13}$ and $\delta$), one with $\theta_{23} < \pi/4$ and one with $\theta_{23} > \pi/4$ (the $(\theta_{23}, \pi/2 - \theta_{23})$ ambiguity). Thus, in general, there can be as much as an eight–fold degeneracy of solutions from a measurement of $P$ and $\bar{P}$ at one $L$ and $E_\nu$ [17], which has already required two experimental runs (one with neutrinos and one with antineutrinos). Further measurements at different oscillation arguments $\Delta$ are needed to eliminate all of the ambiguities, requiring additional runs.

A judicious choice of the $L$ and $E_\nu$ can help reduce the effect of the ambiguities. For example, the $(\delta, \theta_{13})$ ambiguity can be reduced to a $(\delta, \pi - \delta)$ ambiguity by choosing $L/E_\nu = (2n - 1)(410 \text{ km/GeV})(3 \times 10^{-3} \text{ eV}^2/|\delta m_{31}^2|)$, where $n$ is an integer (which implies $\Delta = (n - \frac{1}{2})\pi$, the approximate location of the appearance oscillation peak). The parameter $\theta_{13}$ is removed from the ambiguity (and therefore in principle determined) since the ellipse in $(P, \bar{P})$ space collapses to a line and the lines for different $\theta_{13}$ no longer intersect [16–18]. Furthermore, if $L$ is sufficiently large, the large matter effects separate the ellipses for $\delta m_{31}^2 > 0$ and $\delta m_{31}^2 < 0$, so that they do not intersect, removing the $\text{sgn}(\delta m_{31}^2)$ ambiguity [16,17]. However, measuring $(P, \bar{P})$ at a single $L$ and $E_\nu$ will still leave a $(\delta, \pi - \delta)$ ambiguity, and may have a $\theta_{23}$ ambiguity if $\theta_{23} \neq \pi/4$ (with a possible associated $CPV/CPC$ confusion), so that additional measurements are needed to completely determine the parameters and resolve all of the ambiguities.

The arguments above can be generalized to show that *any* two measurements of neutrino and/or antineutrino appearance probabilities, even if they are not at the same $L$ and $E_\nu$, are subject to the same degeneracy problem. Each of the neutrino or antineutrino probabilities in an appearance experiment (e.g., Eqs. (2) and (3), or (12) and (13)) is a linear combination of the parameters $x^2$, $x \cos \delta$, and $x \sin \delta$. The coefficients of these parameters involve $f$, $\bar{f}$, $g$, and $\Delta$, which depend only on $L$, $E_\nu$, $\delta m_{31}^2$, and $y$. It is not hard to show that for two such measurements of $P$ and/or $\bar{P}$, if $\theta_{13}$ is fixed and $\delta$ is varied, an ellipse is traced out in the two–dimensional probability space. Ellipses with somewhat different values of $\theta_{13}$ will overlap, so that there is always a $(\delta, \theta_{13})$ ambiguity for any two appearance measurements. The only exception is the special situation where the ellipse collapses to a straight line, in which case lines for different $\theta_{13}$ no longer overlap, but there is still an ambiguity involving $\delta$ since the two halves of the collapsed ellipse then overlap (e.g., the $(\delta, \pi - \delta)$ ambiguity for $(P, \bar{P})$ when $\Delta = \pi/2$). Similarly, there is a $\theta_{23}$ ambiguity if $\theta_{23} \neq \pi/4$, and there may be a $\text{sgn}(\delta m_{31}^2)$ ambiguity if the matter effect is not large enough to separate the ellipses for $\delta m_{31}^2 > 0$ and $\delta m_{31}^2 < 0$. Thus two measurements will *always* have an ambiguity involving $\delta$ (which may or may not also involve $\theta_{13}$, depending on the value of the oscillation argument $\Delta$), and may also have ambiguities involving $\text{sgn}(\delta m_{31}^2)$ and $\theta_{23}$. Ambiguities involving $\text{sgn}(\delta m_{31}^2)$ and $\theta_{23}$ will have a $CPV/CPC$ confusion; the ambiguity that involves $\delta$ alone will have a $CPV/CPC$ confusion unless $\Delta = n\pi/2$.

We emphasize that the degeneracies discussed here are exact, i.e., there are different sets of parameters that give *identical* predictions. Thus these ambiguities are present even in the limit of no experimental uncertainties. When such uncertainties are taken into account, nearly degenerate solutions may also be indistinguishable, which would require further (or improved) measurements. Therefore the conditions needed to remove the exact degeneracies are necessary but not sufficient for removing degeneracies in an actual experiment. Furthermore, parameter degeneracies occur nearly everywhere in the $(\delta, \theta_{13})$ plane, since in most cases any given ellipse (for fixed $\theta_{13}$) overlaps an adjacent ellipse (with different $\theta_{13}$).

TABLE I. Four classes of degenerate solutions which could potentially remain after three appearance measurements.

| Case | sgn($\delta m^2_{31}$) | $\theta_{23}$ |
|---|---|---|
| I | + | $< \pi/4$ |
| II | − | $< \pi/4$ |
| III | + | $> \pi/4$ |
| IV | − | $> \pi/4$ |

### B. Parameter degeneracies with three measurements

Since two measurements will generally have parameter degeneracies, the natural question is whether a third measurement sufficient to remove all of them. To extend the degeneracy analysis to more than two appearance measurements, what is needed is a straightforward and systematic method for determining whether or not a given set of such measurements resolves all of the degeneracies over the entire region of interest in the parameters $\delta$ and $\theta_{13}$. The method we present here uses the approximate expressions for the probabilities given in Sec. II.

We begin by analyzing the case of three appearance measurements. From the approximate probability expressions we have three linear equations involving $x^2$, $x\cos\delta$, and $x\sin\delta$. As long as the three equations are linearly independent (which can be realized if no two measurements simultaneously have the same $L$, $E_\nu$, and type of neutrino) these are easily solved, giving a unique solution for $x$ and $\delta$. However, in principle four possible solutions may still exist, corresponding to the sgn($\delta m^2_{31}$) and $\theta_{23}$ ambiguities (see Table I); each such solution gives identical predictions for the three probabilities being measured, but can have different values for $\theta_{13}$ and/or $\delta$.

To determine if the remaining degeneracies have been resolved by the third measurement, we use the fact that the variables $x^2$, $x\cos\delta$, and $x\sin\delta$ are not independent, and in the "true" solution they *must* obey the relation

$$x^2 = (x\cos\delta)^2 + (x\sin\delta)^2 . \tag{16}$$

The question is then whether the "fake" solutions can also satisfy the constraint in Eq. (16). We have found that for many values of $\delta$ and $\theta_{13}$, the fake solutions do not obey Eq. (16) (thus removing the degeneracy). However, for some particular values of $\delta$ and $\theta_{13}$ there exist fake solutions that also obey Eq. (16), so that degeneracy may remain even after three measurements; the values of $\delta$ and $\theta_{13}$ for which degenerate solutions still exist form lines in the ($\delta, \theta_{13}$) plane. It is easy to understand why this is so: with two measurements, most points in the ($\delta, \theta_{13}$) plane have degeneracies, but the condition in Eq. (16) imposes an additional constraint so that for three measurements the dimension of the degenerate space is reduced from two (the plane) to one (lines). Since the forms of $P$ and $\bar{P}$ are the same (i.e., a linear combination of $x^2$, $x\cos\delta$, and $x\sin\delta$), *any* combination of three neutrino and/or antineutrino appearance measurements will yield lines in ($\delta, \theta_{13}$) space where the fake degenerate solutions are not eliminated by three measurements.

For example, consider neutrino measurements made at $L = 730$ km and $E_\nu = 2.66$, 1.77, and 1.34 GeV ($\Delta = \pi/3$, $\pi/2$, and $2\pi/3$, respectively); Fig. 2a shows the lines in $(\delta, \theta_{13})$ space where a fake solution remains degenerate with the true solution (taken to be Case I in the figure). Figure 2b shows similar curves for $L = 730$ km and $E_\nu = 3.54$, 1.77, and 0.89 GeV ($\Delta = \pi/4$, $\pi/2$, and $\pi$, respectively). In Fig. 2 we assume

$$|\delta m^2_{31}| = 3 \times 10^{-3} \text{ eV}^2, \sin^2 2\theta_{23} = 0.90,$$
$$\delta m^2_{21} = 5 \times 10^{-5} \text{ eV}^2, \sin^2 2\theta_{12} = 0.80, \tag{17}$$

which allows the possibility of a $\theta_{23}$ ambiguity as well as the $(\delta, \theta_{13})$ and $\text{sgn}(\delta m^2_{31})$ ambiguities; this will be the standard parameter set used in this paper unless stated otherwise. The key to determining if three measurements are sufficient to remove degeneracies in a particular experiment is whether or not the experiment is sensitive to the values of $\delta$ and $\theta_{13}$ where the lines of degeneracies occur. We see that in the two examples in Fig. 2, degeneracies remain for a wide range of $\delta$ when $\sin^2 2\theta_{13} > 0.01$. Thus either of these two sets of measurements are not sufficient to ensure there are no degeneracies in the next generation of long–baseline neutrino experiments, which should be sensitive down to $\sin^2 2\theta_{13} = 0.01$ or below [19,22,24,30].

## C. Parameter degeneracies with four measurements

When three measurements are not sufficient to completely resolve the degeneracies, a fourth measurement must be made. A straightforward way to determine whether or not the fourth measurement has removed the remaining degeneracies is as follows. Take all possible combinations of three measurements within the four (there are four such combinations) and perform the above analysis on each combination of three measurements. Each three–measurement subgroup may have values of $\delta$ and $\theta_{13}$ for which degeneracies remain (if not, then that subgroup alone is sufficient to remove the degeneracies, and a fourth measurement is not needed). If the degenerate lines in $(\delta, \theta_{13})$ space for the four subgroups do not intersect at a common point, then in principle the four measurements have resolved all the degeneracies; if not, then some degeneracies remain.

For example, in the three measurements used to obtain Fig. 2a there were degeneracies for a range of $\delta$ when $\sin^2 2\theta_{13} > 0.01$. After a fourth measurement with $L = 730$ km and $E_\nu = 5.31$ GeV ($\Delta = \pi/6$) is added to the measurements used in Fig. 2a, the lines of degeneracy for the four subgroups of three measurements are shown in Fig. 3. There is a point near $\sin^2 2\theta_{13} = 0.0085$ and $\delta = 0.19\pi$ where the lines of degeneracy all intersect when comparing I and IV, so that these two cases remain degenerate after these four measurements (see Fig. 3c), Furthermore, there is also a point near $\sin^2 2\theta_{13} = 0.0015$ and $\delta = 0.49\pi$ where the four lines nearly intersect at the same point when comparing I and II, so that these two cases would be degenerate once experimental uncertainties are considered.

It is not hard to understand why some degenerate points remain in $(\delta, \theta_{13})$ space after four measurements. The fourth data point adds one additional constraint on the parameters, which then reduces the dimension of the space that is degenerate from one (lines) to zero (points). For an experiment to be assured of resolving the parameter degeneracies, these degenerate points must lie at $\sin^2 2\theta_{13}$ below the experimentally accessible region.

We note that to determine the parameter degeneracies between, say, Case I and Case II with more than two measurements using a purely numerical approach, for each point in $(\delta, \theta_{13})$ space for Case I, one would have to search the entire $(\delta, \theta_{13})$ space for Case II to check for an identical set of probabilities. With our analytic approach, one need only search the parameter space once, since the existence of parameter degeneracies is determined algebraically.

Figure 4a shows some typical parameter degeneracies between Cases I and II if only one measurement is made for $\nu$ and $\bar{\nu}$ at $L = 300$ km and $\Delta = \pi/2$ (corresponding to $E_\nu = 0.73$ GeV), assuming the same oscillation parameters as Eq. (17) except $\theta_{23} = \pi/4$. We note that these degeneracies can mix CPV and CPC solutions, especially when $\theta_{13}$ is not small; for some sets of parameters, $CPC$ can give identical results to maximal $CPV$ with the opposite $\text{sgn}(\delta m_{31}^2)$. To illustrate how additional measurements reduce the possibilities for degeneracies, Fig. 4b shows the parameter regions in Case I that have degeneracy with Case II, for two, three, and four measurements; the degenerate space is reduced from an area to lines to points. The corresponding degenerate parameters for Case II in Fig. 4b occur at the same $\theta_{13}$ with $\delta \to \delta \pm \pi$; it is not hard to show that this symmetrical situation is an artifact of choosing the same energy for $\nu$ and $\bar{\nu}$, and would not necessarily be true if the $\nu$ and $\bar{\nu}$ energies were different.

In summary, two measurements necessarily have parameter degeneracies over a broad range of $(\delta, \theta_{13})$ space; adding a third measurement reduces the degeneracy to lines in the parameter space, and adding a fourth measurement reduces it further to isolated points. This behavior is to be expected since each additional measurement adds one more constraint, which has the effect of reducing the dimension of the degenerate subspace by one. Adding a fifth measurement with a linearly independent combination of $x^2$, $x \cos \delta$ and $x \sin \delta$ should then remove the degeneracy completely for all values of $\delta$ and $\theta_{13}$. For example, in Fig. 3, if a fifth measurement at $L = 730$ km is made at, say, $\Delta = 5\pi/6$, then no point in the parameter space remains degenerate. Of course, practically speaking, less than five measurements may be sufficient to eliminate the possibility of parameter degeneracies in any given experiment if the degeneracies occur in a region of parameter space in which the experiment is not sensitive. In the next section we explore some specific cases with less than five measurements.

## IV. DETECTOR SCENARIOS FOR RESOLVING PARAMETER DEGENERACIES

In the previous section we presented a general method for determining if a set of more than two measurements are sufficient to remove all of the parameter degeneracies. In this section we discuss specific detector scenarios using this method. We emphasize scenarios where at least some of the measurements can be made at the same $L$ but different $E_\nu$, since these are well–suited to off–axis beams (where different energies can be obtained by having a detector at different locations with respect to the beam axis). If a detector cluster could be used, the time required to make all of the necessary measurements could be reduced. We also compare the detector cluster scenarios to scenarios in which a detector cluster is not used.

## A. Two $\nu$ and two $\bar{\nu}$ measurements at one $L$

The standard means for the study of $CP$ violation is to measure both $\nu_\mu \to \nu_e$ and $\bar{\nu}_\mu \to \bar{\nu}_e$ at the same $L$ and $E_\nu$. As we have seen, there could be as much as an eight–fold degeneracy in the parameters with just these two measurements, and it may be impossible to establish $CPV$ from $\nu$ and $\bar{\nu}$ measurements at a single $L$ and $E_\nu$. The simplest way one could imagine obtaining another set of independent measurements is to measure $\nu_\mu \to \nu_e$ and/or $\bar{\nu}_\mu \to \bar{\nu}_e$ at the same $L$ but with different $E_\nu$; the detector and beamline remain the same, and only the beam energy and/or primary particle in the beam ($\nu_\mu$ or $\bar{\nu}_\mu$) would need to be changed from one run to the next. With a single detector this would take four separate runs. However, with two detectors at different locations with respect to the beam axis (so that each receives neutrinos of different energy) these four measurements could be taken with only two runs (one with $\nu$ and one with $\bar{\nu}$).

Figure 5 shows the points in ($\delta, \theta_{13}$) space where degeneracies remain due to the sgn($\delta m^2_{31}$) and $\theta_{23}$ ambiguities after measurements at two energies for both neutrinos and antineutrinos, assuming that $\bar{\nu}$ energies are the same as the $\nu$ energies and that the measurements are all made at approximately the same $L$, for different choices of $L$. We examined scenarios where one energy was chosen so that $\Delta_1 = \pi/2$, and the second energy was chosen so that $\Delta_2 = \pi/6, \pi/4, \pi/3, 2\pi/3$, or $\pi$; some representative examples are shown in the figure. We chose $\Delta_1 = \pi/2$ because the probabilities are large there (see $f$ and $\bar{f}$ in Fig. 1), the $\cos \delta$ term in the probabilities vanishes (thereby isolating the $\sin \delta$ term), and the survival channel $\nu_\mu \to \nu_\mu$ is near an oscillation minimum. All possible degeneracies are included in Fig. 5 (e.g., I vs. II, II vs. I, I vs. III, etc., where the first case is the true solution), and different types of degeneracies are not distinguished. In all cases there are some points in the parameter space where degeneracies are still possible. We found that if $\Delta_2 < \pi/2$, the degeneracies tended to be at smaller values of $\theta_{13}$ than for $\Delta_2 > \pi/2$; the probabilities are also higher there (see $f$ and $\bar{f}$ in Fig. 1).

For off–axis beams, both the neutrino flux and energy fall off with increasing off–axis angle (see Eqs. 1 and 2 of Ref. [23]); after eliminating the angular dependence one obtains for the energy dependence of the flux

$$\Phi \propto E_\nu^2, \tag{18}$$

so that $\Phi \propto \Delta^{-2}$ for measurements at the same $L$, Therefore if two off–axis measurements had very different $\Delta$ values, one would have a much reduced flux compared to the other. Hence it may be preferable if the ratio between the two values of $\Delta$ is not too large, since then the fluxes would be more equal. For example, with $\Delta_1 = \pi/2$, $\Delta_2 = \pi/4$ and $\pi/3$ give very similar parameter degeneracies, but choosing $\Delta_2 = \pi/3$ would make the fluxes of the two measurements closer; even then, the ratio of the fluxes would be 4:9. Of course, the detector further off axis could be made larger to help equalize the event rate. The detectors should not be too close in off–axis angle, either, since then the $\Delta$ values would be very similar and the expressions for the probabilities would no longer be linearly independent after experimental uncertainties are taken into account.

Similar to the discussion of Fig. 4b, since the $\bar{\nu}$ energies are the same as the $\nu$ energies, the degenerate points in Fig. 5 come in symmetrical pairs, each with the same $\theta_{13}$ and $\delta$ differing by $\pi$. It can be shown that the degenerate parameters for Case I (III) when it is degenerate with Case II (IV) exhibit this symmetry property with the parameters for II (IV) when it is degenerate with I (III). Therefore for degeneracies between I and II, or between III and IV, $CPC$ and $CPV$ solutions are not mixed, since $\delta$ and $\delta \pm \pi$ are either both $CPC$ or $CPV$ (although of course sgn($\delta m^2_{31}$) is not determined). On the other hand, the symmetry also exists between the parameters for I when it is degenerate with IV (III) and the parameters for II when it is degenerate with III (IV), and between the parameters for IV (III) when it is degenerate with I and the parameters for III (IV) when it is degenerate with II; in all of these situations, the degenerate points that exhibit the symmetry property do not occur between two cases that are degenerate with each other, so there is still a $CPC/CPV$ confusion for these degeneracies.

The values of $\delta$ at which degeneracies occur in Fig. 5 are insensitive to the assumed values of $\delta m^2_{21}$ and $\sin^2 2\theta_{12}$. However, we found that the values of $\theta_{13}$ at which degeneracies occur vary approximately as

$$\sin^2 2\theta_{13}^{degen} = (\sin^2 2\theta_{13}^{degen})_0 \left( \frac{\delta m^2_{21}}{5 \times 10^{-5} \text{ eV}^2} \right)^2 \left( \frac{\sin^2 2\theta_{12}}{0.80} \right), \qquad (19)$$

where the degenerate parameter is $(\sin^2 2\theta_{13}^{degen})_0$ when $\delta m^2_{21} = 5 \times 10^{-5}$ eV$^2$ and $\sin^2 2\theta_{12} = 0.80$. Thus the values of $\theta_{13}$ at which degeneracies occur is somewhat sensitive to $\sin^2 2\theta_{12}$ (which varies from 0.67 to 0.92 at the $2\sigma$ level [4]) and very sensitive to $\delta m^2_{21}$ (which varies from $3 \times 10^{-5}$ to $2 \times 10^{-4}$ eV$^2$ at the $2\sigma$ level [4]). We also found that a variation in $\delta m^2_{31}$ of 10% (the expected precision of MINOS, ICARUS, and OPERA) caused $\sin^2 2\theta_{13}$ of the degenerate points to vary by 30% or less; smaller $\delta m^2_{31}$ moved the degeneracies to higher $\sin^2 2\theta_{13}$; the largest variations occurred at lower $\sin^2 2\theta_{13}$. The corresponding variation in $\delta$ was much less, of order a few percent.

In Fig. 5, all of the examples at shorter $L$ have degeneracies for $\sin^2 2\theta_{13} \simeq 0.01$ or larger, which is within the expected range of future long–baseline experiments with superbeams. However, at $L = 300$ km with $\Delta = \pi$, the degeneracy for larger $\theta_{13}$ occurs for $\sin^2 2\theta_{13} > 0.1$ (around 0.13), above the current upper bound. The only degeneracies with $\sin^2 2\theta_{13} < 0.1$ in the $L = 300$ km case occur at much smaller $\theta_{13}$ ($\sin^2 2\theta_{13} \simeq 0.0013$; see Fig. 5c), just below the region accessible at the $3\sigma$ level in the SuperJHF to Hyper–K experiment [22]. Thus $\nu$ and $\bar{\nu}$ measurements with both $\Delta = \pi/2$ and $\pi$ at $L = 300$ km do not have degeneracy problems over the entire search region. However, the values of $|f|$ and $|\bar{f}|$ are significantly smaller for $\Delta = \pi$ (by more than a factor of 10 compared to $\Delta = \pi/2$; see Fig. 1), so that the probabilities are also significantly smaller there, leading to significantly reduced statistics that will limit the $\sin^2 2\theta_{13}$ reach of such an experiment. Furthermore, for slightly smaller values of $\delta m^2_{21}$ or $\sin^2 2\theta_{12}$ the degeneracy at $\sin^2 2\theta_{13} = 0.13$ will be pushed below 0.10, into the search region.

Since the matter effect increases with increasing $L$ and $\theta_{13}$, one would expect that at larger $L$ degenerate solutions with opposite sgn($\delta m^2_{31}$) can be differentiated for smaller values of $\theta_{13}$. This property is evident in Fig. 5; e.g., for $L = 2900$ km degeneracies occur only at $\sin^2 2\theta_{13} \lesssim 0.002$–0.003. If such an experiment were sensitive only to $\sin^2 2\theta_{13}$ above this value, it would not have a degeneracy problem. The $\sin^2 2\theta_{13}$ sensitivity depends on the total flux (which varies with $L$ and off–axis angle), so a detailed calculation must be made to determine the existence of degeneracies in any given experimental situation.

There may also be approximate degeneracies that cannot be differentiated due to statistical and systematic uncertainties. Here, too, one also needs to include details of a particular experimental situation to determine if approximate degeneracies are a problem; such an analysis is beyond the scope of this paper. However, to give an indication of the effect of approximate degeneracies, in Fig. 6 we show regions in $(\delta, \theta_{13})$ space where the differences between $(x\cos\delta)^2 + (x\sin\delta)^2$ and $x^2$ for the false solution, when added in quadrature, are less than 10%. We see that in Figs. 6a and 6b the approximate regions extend to $\sin^2 2\theta_{13}$ that are nearly twice as large as the maximum values for the exact degeneracies. Furthermore, in Fig. 6c there are no exact degeneracies in the region shown, but approximate degeneracies extend to $\sin^2 2\theta_{13}$ as large as 0.007. Therefore it is clear that approximate degeneracies will occur at somewhat higher $\theta_{13}$ than the exact degeneracies.

### B. Four $\nu$ measurements at one $L$

Antineutrino measurements have two drawbacks compared to neutrino measurements: the $\bar{\nu}$ cross section is approximately half as large, and $\bar{\nu}$ beam fluxes are less. Thus to achieve comparable statistics the running time for an antineutrino measurement must be more than twice as long as a similar neutrino measurement (apart from the possible enhancement of antineutrino oscillation probabilities at long $L$ if $\delta m_{31}^2 < 0$). If it were possible to remove the degeneracies with neutrino measurements alone, the total time to complete all of the necessary measurements would be substantially reduced. With a single detector, this would require more than two separate runs (since, as we have shown, any two measurements alone will have degeneracies), each at a different $E_\nu$. With two detectors at different locations with respect to the beam axis, only two runs would give four independent measurements, and only the on–axis beam energy or the detector off-axis angles would have to be changed between runs (although the latter might prove to be unfeasible). Another interesting possibility is to have a four-detector cluster, so that only a single neutrino run would be needed to eliminate all degeneracies. Although this many detectors might seem to be economically unfeasible, the running time is shorter and the beam would not have to be reconfigured, which would offset increased detector costs. In this section we examine the ability of four $\nu$ runs to resolve degeneracies.

In Fig. 7 we show the points in the $(\delta, \theta_{13})$ plane that have degeneracy with four neutrino measurements at the same $L$, for several choices of $L$. We see that in all cases there are points in the parameter space that have degeneracies even after four measurements are taken. Similar to the case with two $\nu$ and two $\bar{\nu}$ measurements, the degenerate solutions are pushed to lower $\theta_{13}$ as $L$ is increased, due to the larger matter effect, which helps distinguish $\text{sgn}(\delta m_{31}^2)$. For $L = 2900$ km, in the best cases we found that the largest $\sin^2 2\theta_{13}$ with a degeneracy is 0.004–0.006. A comparison of Figs. 5 and 7 shows that two $\nu$ and two $\bar{\nu}$ measurements at the same $L$ generally has degeneracies occurring at lower $\sin^2 2\theta_{13}$ than four $\nu$ measurements at one $L$.

For a small off–axis angle $\theta$, the neutrino energy is given approximately by [23]

$$E_\nu \simeq \frac{0.43 E_\pi}{1 + (E_\pi \theta/m_\pi)^2}, \tag{20}$$

where $E_\pi$ is the pion energy in the lab frame and $m_\pi$ is the pion mass. Four different measurements can be made simultaneously if four detectors are placed at different off–axis angles, and it is always possible to choose $E_\pi$ and the angles to select any desired combination of $\Delta_i$. This is not the case if only two detectors are used and two runs are made with different $E_\pi$ (assuming the detectors stay at the same off–axis angles in both runs). From Eq. (20), the condition that a set of four $\Delta_i$ can be obtained in a two–detector, two–run scenario is

$$\Delta_1(\Delta_4 - \Delta_1) \geq \Delta_2(\Delta_3 - \Delta_2), \tag{21}$$

where $\Delta_4 > \Delta_3 > \Delta_2 > \Delta_1$. In practice, Eq. (21) is not hard to satisfy, especially if $\Delta_4$ is not too small and the difference between $\Delta_3$ and $\Delta_2$ is not too large. All of our examples in Fig. 7 can be obtained in two runs with two detectors.

One drawback of making only neutrino measurements is that this is not a direct measurement of $CPV$ (such as would be the case of having both $\nu$ and $\bar\nu$ measurements). Furthermore, if $\delta m^2_{31} < 0$, then all of the measurements would have a suppression due to matter effects, especially at longer $L$. A detailed calculation including experimental uncertainties would have to be made to see if the improved resolution of degeneracies at longer $L$ can overcome the rate loss if $\delta m^2_{31} < 0$.

### C. Measurements at more than one $L$

In the previous examples all of the detectors were assumed to be at the same distance. As we have seen, measurements at longer $L$ push the degeneracies to lower $\sin^2 2\theta_{13}$. Here we consider having two $\nu$ measurements made at one distance (simultaneously with a detector cluster) and a third $\nu$ measurement at another distance. We examine two possibilities: one with the detector cluster at the shorter distance, the other with the detector cluster at the longer distance.

In all cases we assume the first measurement is made with a $\nu$ beam at $L_1 = 300$ km and $\Delta_1 = \pi/2$ (the approximate values for the proposed JHF to Super–K experiment). If the second $\nu$ measurement is also at $L_2 = 300$ km with $\Delta_2 \neq \Delta_1$, and a third $\nu$ measurement at $L_3 > L_2$, we found that the degeneracies are less severe (i.e., they occur at smaller $\sin^2 2\theta_{13}$) when $\Delta_3 = \Delta_2 < \pi/2$; some of the better examples we found are shown in Table II. For the longer distance we used the $L$ values considered in Fig. 7, plus 2100 km (nominally JHF to Beijing). Having $\Delta_2 < \pi/2$ and $\Delta_3 = \pi/2$ also gave good results. In the cases we tested with the second and third $\nu$ measurements both at longer $L$, the best results were achieved if one of the measurements at the longer distance had $\Delta = \pi/2$; some of the better examples we found are shown in Table III.

In both Tables II and III, experimental $L$ and $E_\nu$ choices can be made such that degeneracies occur only at $\sin^2 2\theta_{13} = 0.003$ or less (assuming standard values for $\sin^2 2\theta_{12}$ and $\delta m^2_{12}$). Thus, although there are only three measurements, with appropriate choices of $L$ and $\Delta$, degeneracies are less of a problem (i.e., they occur at smaller $\theta_{13}$) than for the scenarios where all measurements are done at the same $L$. Of course, for only three measurements the regions where degeneracies occur are lines in $(\delta, \theta_{13})$ space, rather than isolated points, so they are more likely to occur in experiments with $\sin^2 2\theta_{13}$ sensitivity of order 0.003 or less. Furthermore, as with the scenario with four $\nu$ measurements at one $L$, a determination of $CPV$ is indirect (depending on the three–neutrino parametrization) and probabilities would be suppressed if $\delta m^2_{31} < 0$.

TABLE II. Maximum $\sin^2 2\theta_{13}$ that has a degeneracy when the first $\nu$ measurement is made at $L_1 = 300$ km and $\Delta_1 = \pi/2$, a second at $L_2 = 300$ km with $\Delta_2 \neq \Delta_1$, and a third at $L_3 > L_2$, for different choices of the second and third measurements. The other neutrino parameters are the same as in Eq. (17).

| $\Delta_2(\nu)$ | $\Delta_3(\nu)$ | $L_3$ | | | | |
|---|---|---|---|---|---|---|
| | | 730 km | 1290 km | 1770 km | 2100 km | 2900 km |
| $\pi/4$ | $\pi/4$ | .0012 | .0013 | .0013 | .0013 | .0013 |
| | $\pi/2$ | .0028 | .0033 | .0038 | .0042 | .0060 |
| $\pi/3$ | $\pi/3$ | .0011 | .0012 | .0012 | .0013 | .0013 |
| | $\pi/2$ | .0021 | .0025 | .0030 | .0034 | .0047 |

TABLE III. Maximum $\sin^2 2\theta_{13}$ that has a degeneracy when the first $\nu$ measurement is made at $L_1 = 300$ km and $\Delta_1 = \pi/2$, and two other $\nu$ measurements are made at $L_2 = L_3 > L_1$, for different choices of $\Delta$ for the second and third measurements. The other neutrino parameters are the same as in Eq. (17).

| $\Delta_2(\nu)$ | $\Delta_3(\nu)$ | $L_2 = L_3$ | | | | |
|---|---|---|---|---|---|---|
| | | 730 km | 1290 km | 1770 km | 2100 km | 2900 km |
| $\pi/2$ | $\pi/4$ | .0027 | .0030 | .0033 | .0038 | .0047 |
| $\pi/2$ | $\pi/3$ | .0021 | .0022 | .0025 | .0027 | .0035 |
| $\pi/2$ | $2\pi/3$ | .0030 | .0042 | .0067 | .0094 | .032 |

### D. Scenarios without a detector cluster

All of our previous examples included some measurements at more than one energy, but at the same $L$, which is particularly suited to detector cluster experiments. Here we examine scenarios which do not use a detector cluster.

First we consider a $\nu$ and $\bar{\nu}$ measurement at one $L$ and $E_\nu$, and another $\nu$ measurement at the same $L$ and different $E_\nu$. This is similar to the case of two $\nu$ and two $\bar{\nu}$ measurements all at the same $L$ discussed in Sec. IV A, except that the second $\bar{\nu}$ measurement is missing, and the two $\nu$ measurements are done in separate runs with different $\nu$ energies. Since there are only three measurements, the parameter degeneracies are lines in $(\delta, \theta_{13})$ space. We examined cases where $\nu$ and $\bar{\nu}$ measurements were done at $\Delta = \pi/2$ and another $\nu$ measurement was done at $\Delta = \pi/6$, $\pi/4$, $\pi/3$, $2\pi/3$, $3\pi/4$, or $\pi$, for $L = 300$, 730, 1290, 1770, 2100, or 2900 km. We found that for $L \leq 1290$ km, in the best cases degeneracies can occur for $\sin^2 2\theta_{13}$ as high as 0.01 (0.10 for $L = 300$ km). For the best cases with $L \geq 1770$ km degeneracies can occur for $\sin^2 2\theta_{13}$ as high as 0.005-0.008, which is a factor of two or more higher than for two $\nu$ and two $\bar{\nu}$ energies at the same $L$. Thus making a second $\bar{\nu}$ measurement not only reduces the set of degenerate points from lines to points, but it also pushes the degeneracies to lower $\sin^2 2\theta_{13}$.

Next we consider a $\nu$ and $\bar{\nu}$ measurement at one $L$ and $E_\nu$, and another $\nu$ measurement at a different $L$. The neutrino energies for the two $\nu$ measurements may or may not be the same. We examine two possibilities, where the $\bar{\nu}$ measurement is done at either the shorter or longer $L$. In all cases we assume the first measurement is made with a $\nu$ beam at $L_1 = 300$ km and $\Delta_1 = \pi/2$. If the $\bar{\nu}$ measurement is also at $L_2 = 300$ km with $\Delta_2 = \pi/2$, and a $\nu$ measurement at $L_3 > 300$ km, we found that for $\Delta_3 \simeq \pi/3$ ($\Delta_3 \simeq 2\pi/3$) the degeneracies are less severe at larger (intermediate) $L_3$; see Table IV. Another possibility is to have the $\bar{\nu}$ measurement done at the longer $L$. Typical results are shown in Table V.

TABLE IV. Maximum $\sin^2 2\theta_{13}$ that has a degeneracy when $\nu$ and $\bar\nu$ measurements are made at $L_1 = L_2 = 300$ km and $\Delta_1 = \Delta_2 = \pi/2$ and another $\nu$ measurement is made at $L_3 > 300$ km, for different choices of $L_3$ and $\Delta_3$. The other neutrino parameters are the same as in Eq. (17).

| $\Delta_3(\nu)$ | $L_3$ | | | | |
|---|---|---|---|---|---|
| | 730 km | 1290 km | 1770 km | 2100 km | 2900 km |
| $\pi/3$ | .089 | .018 | .008 | .006 | .004 |
| $2\pi/3$ | .035 | .010 | .016 | .025 | .071 |

TABLE V. Maximum $\sin^2 2\theta_{13}$ that has a degeneracy when a $\nu$ measurement is made at $L_1 = 300$ km and $\Delta_1 = \pi/2$, and $\nu$ and $\bar\nu$ measurements are made at $L_2 = L_3 > 300$ km, for different choices of the second and third measurements. The other neutrino parameters are the same as in Eq. (17).

| $\Delta_2\ (\nu)$ | $\Delta_3\ (\bar\nu)$ | $L_2 = L_3$ | | | | |
|---|---|---|---|---|---|---|
| | | 730 km | 1290 km | 1770 km | 2100 km | 2900 km |
| $\pi/3$ | $\pi/2$ | .050 | .014 | .005 | .006 | .010 |
| $2\pi/3$ | $\pi/2$ | .038 | .010 | .016 | .022 | .067 |
| $\pi/4$ | $\pi/4$ | .056 | .013 | .007 | .005 | .008 |
| $\pi/3$ | $\pi/3$ | .038 | .008 | .005 | .006 | .013 |
| $2\pi/3$ | $2\pi/3$ | .033 | .011 | .011 | .015 | .063 |

Generally speaking the measurements in Tables IV and V have degeneracies for higher $\sin^2 2\theta_{13}$ than with three $\nu$ measurements, one at a different $L$ (Tables II and III).

## V. SUMMARY AND DISCUSSION

We summarize the important points of our paper as follows:

(i) For any two appearance measurements there may be as much as an eight–fold parameter degeneracy (resulting from simultaneous $(\delta, \theta_{13})$, $\text{sgn}(\delta m_{31}^2)$ and $(\theta_{23}, \pi/2 - \theta_{23})$ ambiguities) for most points in the $(\delta, \theta_{13})$ plane. Making a third appearance measurement resolves the $(\delta, \theta_{13})$ ambiguity and reduces the regions where the remaining degeneracies occur to lines in $(\delta, \theta_{13})$ space. Making a fourth appearance measurement reduces these degeneracies to isolated points in the $(\delta, \theta_{13})$ plane; a fifth measurement then in principle removes all remaining degeneracies.

(ii) Two $\nu$ and two $\bar\nu$ measurements at the same $L$ can be made so that parameter degeneracies only occur at isolated points in $(\delta, \theta_{13})$ space at $\sin^2 2\theta_{13} \leq 0.01$–$0.02$ ($0.002$–$0.003$) for $L = 300$ ($2900$) km. If the $\bar\nu$ energies are the same as the $\nu$ energies, there will be no $CPC/CPV$ confusion for degeneracies between Cases I and II (defined in Table I), or between Cases III and IV, although $\text{sgn}(\delta m_{31}^2)$ is not determined. This scenario could be completed in two runs (one with a $\nu$ beam and one with a $\bar\nu$ beam) with two detectors at different positions with respect to the beam axis (a two–detector cluster).

(iii) Four $\nu$ measurements at the same $L$ can be made so that parameter degeneracies only occur at isolated points in $(\delta, \theta_{13})$ space at $\sin^2 2\theta_{13} < 0.04$ ($0.004$) for $L = 300$ ($2900$) km. Such a scenario could be implemented in two runs (two $\nu$ beams with different on–axis energies) with a two–detector cluster, or in one run (a single $\nu$ beam) with a four–detector cluster. It would have the advantage that the $\nu$ cross section and flux would be larger than for $\bar\nu$; disadvantages include the fact that it is an indirect measurement of $CPV$ and it would have rates suppressed by matter effects at longer $L$ if $\delta m^2_{31} < 0$.

(iv) Two $\nu$ measurements at one $L$ and a third at another $L$ can be made so that parameter degeneracies only occur at $\sin^2 2\theta_{13} \leq 0.001$–$0.003$. Since there are three measurements, the degeneracies occur along lines in the $(\delta, \theta_{13})$ plane. Such a scenario could be implemented in two runs, with a two–detector cluster at one $L$ and a single detector at the other $L$. This scenario uses the matter effect at longer $L$ to help push degeneracies to lower $\theta_{13}$, and, since it uses only $\nu$ beams, has similar advantages and disadvantages as (iii).

(v) Both a $\nu$ oscillation measurement and a $\bar\nu$ oscillation measurement at one $L$ and a $\nu$ measurement at another $L$ can be made so that parameter degeneracies occur along lines in the $(\delta, \theta_{13})$ plane for $\sin^2 2\theta_{13} \leq 0.005$–$0.010$. Similar results can be obtained for two separate $\nu$ measurements and one $\bar\nu$ measurement at the same $L$. These scenarios do not use a detector cluster.

(vi) All of the examples we show assume $\delta m^2_{21} = 5 \times 10^{-5}$ eV$^2$ and $\sin^2 2\theta_{12} = 0.80$. We found that the maximum $\theta_{13}$ that may have parameter degeneracies varied strongly with $\delta m^2_{21}$, and less so with $\sin^2 2\theta_{12}$ (see Eq. 19); the corresponding $\delta$ values were unaffected. There was also a dependence on $\delta m^2_{31}$, but if $\delta m^2_{31}$ is known to 10% its uncertainty does not greatly affect our degeneracy analysis.

A summary of the requirements for the scenarios discussed in Sec. IV is given in Table VI; also shown is the best–case (i.e., lowest) maximum value of $\sin^2 2\theta_{13}$ that we found for which a parameter degeneracy could occur, assuming the standard parameter set of Eq. (17). Generally speaking, we found that measurements at more than one $L$ did better at resolving parameter degeneracies than if all measurements were done at a single $L$; for example, a third $\nu$ measurement at a different $L$ than the first two measurements appears to be better than a third and a fourth at the same $L$. Also, scenarios with detector clusters resolved degeneracies as well as or better than those without a detector cluster, and required fewer runs. Most scenarios resolved parameter degeneracies the best at longer $L$ (1770 km or more), except for the scenario with a two–detector cluster at one $L$ and a single detector at another $L$, which actually did better when the second distance was not as long (1290 km or less).

If $\theta_{23} \simeq \pi/4$ (the value favored by atmospheric neutrino experiments), then the $\theta_{23}$ ambiguity vanishes and the only two cases to consider after three measurements are I and II (the sgn($\delta m^2_{31}$) ambiguity). We have also performed the degeneracy analyses for $\sin^2 2\theta_{23} = 1$, and found that all of the relevant degeneracies are at least a factor of two lower in $\sin^2 2\theta_{13}$ than shown in Table VI for the shorter distances ($L \leq 1290$ km), and in most cases lower than that for the longer distances. Therefore if it is known that $\theta_{23}$ is close to $\pi/4$ (for example, from measurements at MINOS or ICARUS), the parameter degeneracy problem is greatly reduced.

Some of the scenarios we discussed involve measurements with neutrinos only (not antineutrinos). Although it is possible to extract the neutrino mixing angles and $CP$ phase from such measurements, the result relies on the three–neutrino mixing assumption. A more robust determination of $CP$ violation which would not be as model dependent would include a measurement involving antineutrinos; then the measurement of an asymmetry between neutrino and antineutrino oscillation probabilities (after correcting for the $CP$ violation induced by matter effects) would be direct and definitive.

TABLE VI. A summary of detector scenarios discussed in this paper. The right–most column indicates the best–case (i.e., lowest) maximum value of $\sin^2 2\theta_{13}$ that we found for which a parameter degeneracy could occur, assuming the parameters of Eq. (17).

| # of detectors | # of beams | # of beamlines | # of runs | # of measurements | max. $\sin^2 2\theta_{13}$ without degeneracy |
|---|---|---|---|---|---|
| 2 at one $L$ | 2 ($\nu$ and $\bar{\nu}$) | 1 | 2 | 4 | .002 |
| 2 at one $L$ | 2 ($\nu$ at $E_1$ and $E_2$) | 1 | 2 | 4 | .004 |
| 4 at one $L$ | 1 ($\nu$) | 1 | 1 | 4 | .004 |
| 2 at $L_1$, 1 at $L_2$ | 1 ($\nu$) | 2 | 2 | 3 | .001 |
| 1 at one $L$ | 3 ($\nu$ at $E_1$ and $E_2$, $\bar{\nu}$) | 1 | 3 | 3 | .005 |
| 1 at $L_1$, 1 at $L_2$ | 2 ($\nu$ and $\bar{\nu}$) or 3 (2 $\nu$ and 1 $\bar{\nu}$) | 2 | 3 | 3 | .004 |

The detector cluster scenarios are especially well–suited for detector designs that emphasize large, cheaply–built detectors. We note that the detectors in a cluster would not necessarily have to be the same size; detectors at larger off–axis angles could be made larger to compensate for the reduced flux off–axis.

We emphasize that our analysis only considered exact degeneracies for monoenergetic beams. Experimental uncertainties will expand the regions with degeneracies, especially at small $\sin^2 2\theta_{13}$ (where probabilities and hence events rates are lower) and at longer $L$ (where beam fluxes fall off). On the other hand, neutrino beams are not precisely monoenergetic, and energy spectrum information can help resolve the degeneracies, even for relatively narrow spectrum beams [19]; the survival channel $\nu_\mu \to \nu_\mu$ can in principle also provide additional information, although the effects of $\theta_{13}$ and $\delta$ are not at leading order. A definitive analysis would have to include all of these factors to determine if parameter degeneracies are a problem in a particular experiment. The results of this paper can serve as a guideline for which detector/beam scenarios are likely to encounter difficulties with parameter degeneracies.

## ACKNOWLEDGMENTS

This research was supported in part by the U.S. Department of Energy under Grants No. DE-FG02-95ER40896, No. DE-FG02-01ER41155 and No. DE-FG02-91ER40676, and in part by the University of Wisconsin Research Committee with funds granted by the Wisconsin Alumni Research Foundation.

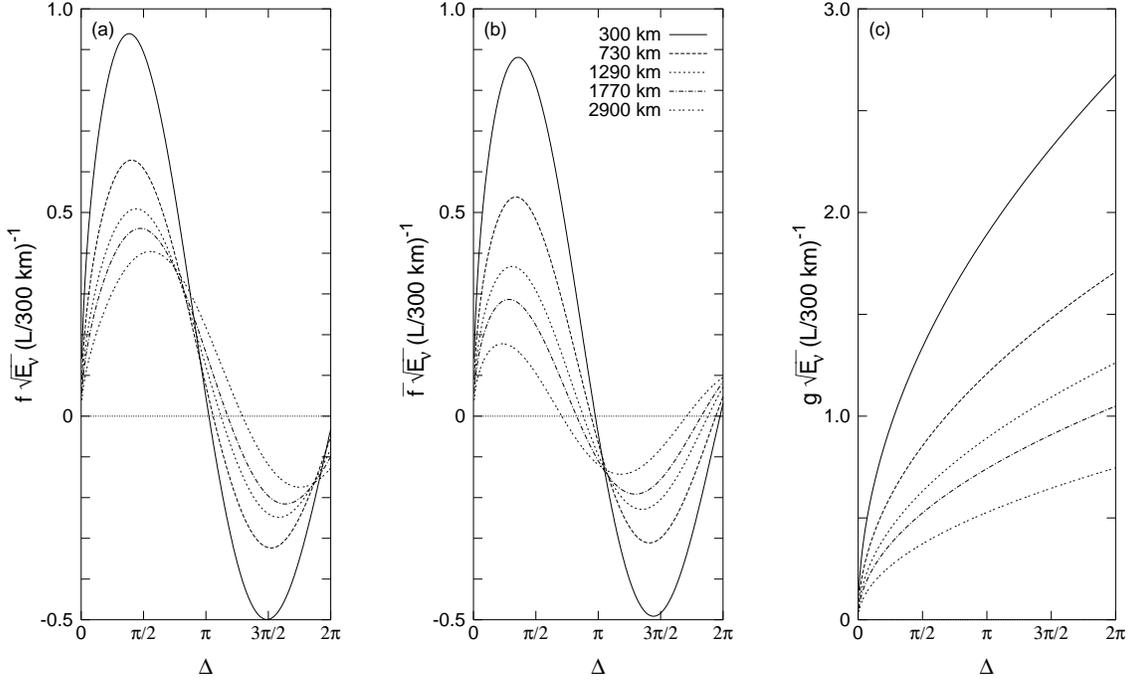

FIG. 1. Value of $\sqrt{E_\nu}/L$ times the coefficients (a) $f$, (b) $\bar{f}$, and (c) $g$ versus $\Delta$ for several values of $L$, assuming $\delta m^2_{31} = 3 \times 10^{-3}$ eV$^2$, and where $E_\nu$ is in GeV.

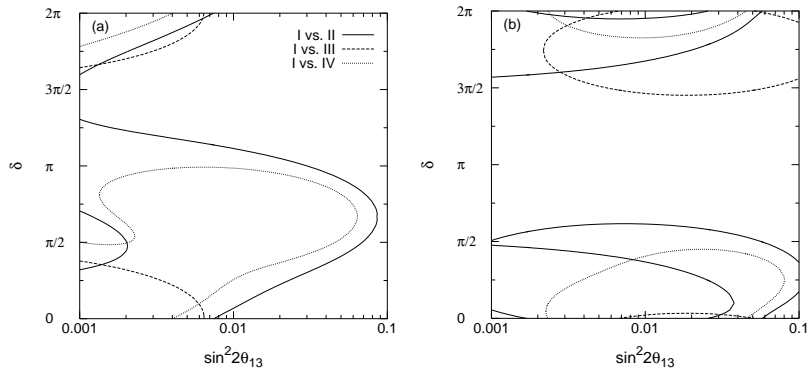

FIG. 2. Lines in $(\sin^2 2\theta_{13}, \delta)$ space for Case I where three neutrino measurements are not sufficient to eliminate the ambiguity between I and II (solid curves), I and III (dashed), and I and IV (dotted), assuming $L = 730$ km and that the other parameters have the values given in Eq. (17). The values of $E_\nu$ are chosen so that the oscillation arguments are (a) $\Delta = \pi/3$, $\pi/2$, and $2\pi/3$, ($E_\nu = 2.66$, 1.77 and 1.33 GeV, respectively), and (b) $\Delta = \pi/4$, $\pi/2$, and $\pi$ ($E_\nu = 3.54$, 1.77 and 0.89 GeV, respectively).

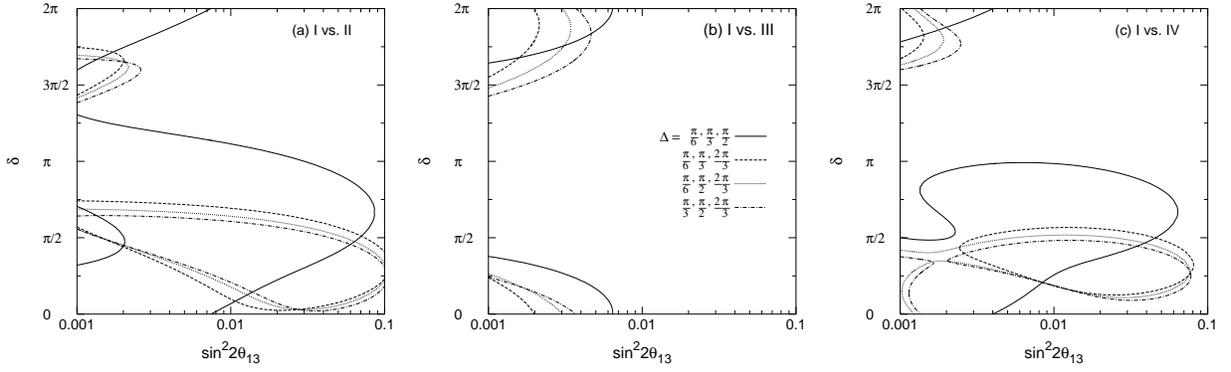

FIG. 3. Lines in $(\sin^2 2\theta_{13}, \delta)$ space for Case I where subgroups of three neutrino measurements are not sufficient to eliminate the parameter degeneracies between (a) I and II, (b) I and III, and (c) I and IV, assuming $L = 730$ km and that the other parameters are the same as in Eq. (17). The values of $E_\nu$ for the four measurements are chosen so that the oscillation arguments are $\Delta = \pi/6$, $\pi/3$, $\pi/2$, and $2\pi/3$, ($E_\nu = 5.31$, 2.66, 1.77 and 1.33 GeV, respectively), and the four subgroups are labeled as follows: $E_\nu = 5.31$ GeV omitted (solid), $E_\nu = 2.66$ GeV omitted (dashed), $E_\nu = 1.77$ GeV omitted (dotted), $E_\nu = 1.33$ GeV omitted (dash–dotted).

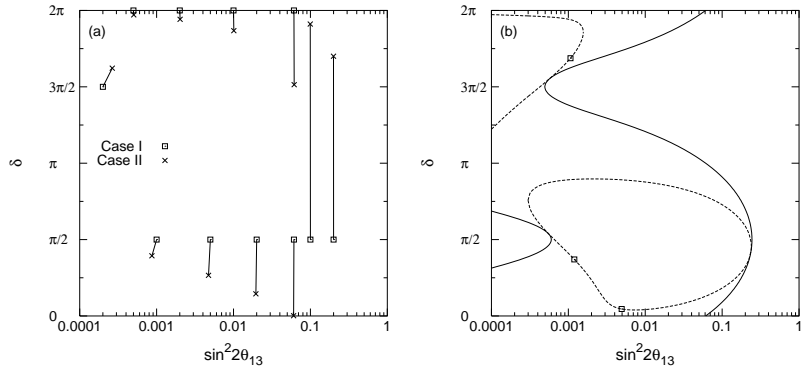

FIG. 4. (a) Sample points in $(\delta, \theta_{13})$ space that have parameter degeneracies between Cases I (boxes) and II (crosses), defined in Table I, after $\nu$ and $\bar{\nu}$ measurements at $L = 300$ km and $\Delta = \pi/2$ (corresponding to $E_\nu = 0.73$ GeV), assuming the other parameters are the same as in Eq. (17), except for $\theta_{23} = \pi/4$. The sets of parameters that are degenerate with each other are linked by a line. For each pair of linked points there is another pair with $\delta \to \pi - \delta$ (not shown). (b) Region in Case I parameter space that have degeneracies with Case II after (i) the two measurements in (a) (area between the solid curves), (ii) an additional $\nu$ measurement at $L = 300$ km and $\Delta = \pi/3$ (along dashed curves), and (iii) a final $\bar{\nu}$ measurement at $L = 300$ km and $\Delta = \pi/3$ (boxes). The corresponding regions for Case II are found by taking $\delta \to \delta \pm \pi$. We note that after only the first two measurements there is also a $(\delta, \pi - \delta)$ ambiguity everywhere in the plane, which is removed by the third measurement.

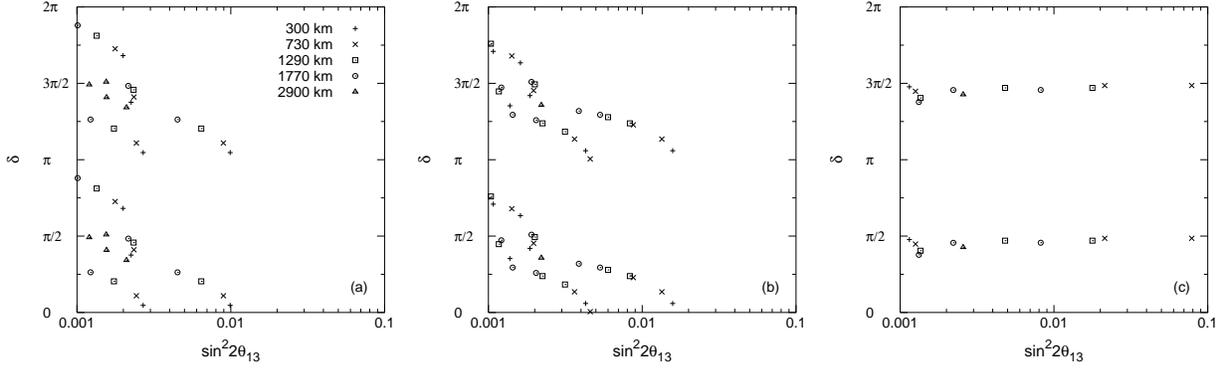

FIG. 5. Points in $(\sin^2 2\theta_{13}, \delta)$ space where two $\nu$ and two $\bar{\nu}$ measurements are not sufficient to eliminate all of the parameter degeneracies, for several values of $L$, assuming the other parameters are the same as in Eq. (17). In each case measurements assumed to be made at two energies such that $\Delta = \pi/2$ and (a) $\Delta = \pi/3$, (b) $\Delta = 2\pi/3$, and (c) $\Delta = \pi$.

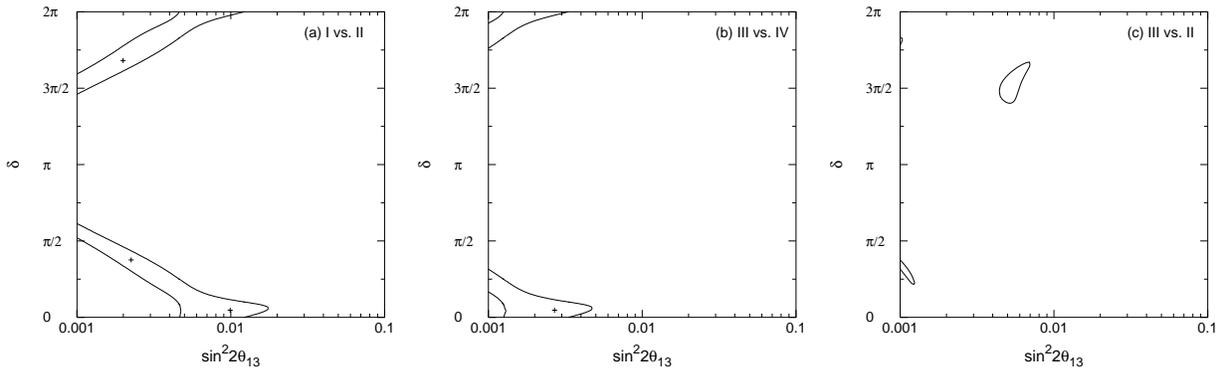

FIG. 6. Regions in $(\sin^2 2\theta_{13}, \delta)$ space where two $\nu$ and two $\bar{\nu}$ measurements at $L = 300$ km with $\Delta = \pi/2$ and $\Delta = \pi/3$ still have approximate parameter degeneracies, assuming the other parameters are the same as in Eq. (17). The three degeneracies shown are (a) I vs. II (I is the true solution and II is the false solution). (b) III vs. IV, and (c) III vs. II. The condition defining the approximate region is given in the text. Exact parameter degeneracies are indicated with pluses.

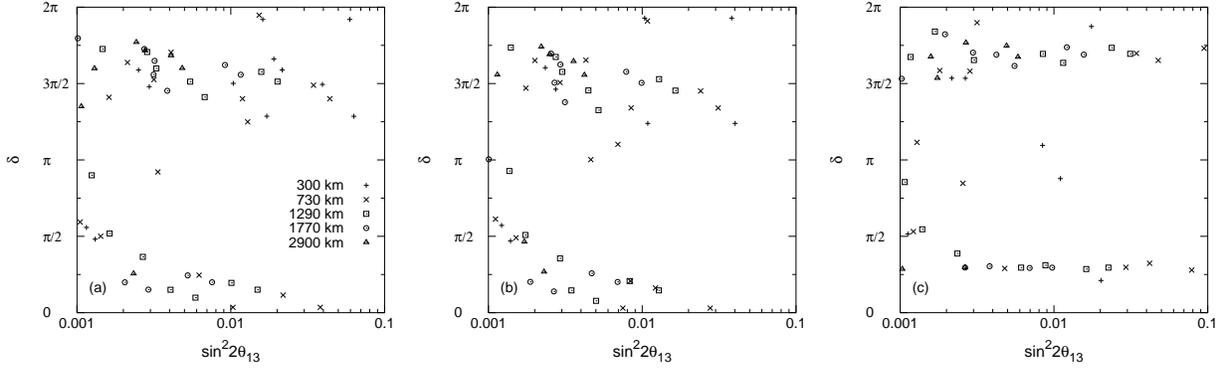

FIG. 7. Points in $(\sin^2 2\theta_{13}, \delta)$ space where four $\nu$ measurements are not sufficient to eliminate all of the parameter degeneracies, for several values of $L$, assuming the other parameters are the same as in Eq. (17). In each case measurements are made at four energies such that (a) $\Delta = \pi/4$, $\pi/3$, $\pi/2$, and $3\pi/4$, (b) $\Delta = \pi/6$, $\pi/3$, $\pi/2$, and $2\pi/3$, and (c) $\Delta = \pi/6$, $\pi/3$, $\pi/2$, and $\pi$.